\documentstyle[psfig]{mn}

\def\nat{Nature}
\def\mnras{MNRAS}
\def\aap{A\&A}
\def\apj{ApJ}
\def\apjl{ApJL}
\def\apjs{APJS}
\def\aj{AJ}
\def\e{{\rm e}}

\title{On the Formation of Cluster Radio Relics}

\author[T. A. En{\ss}lin, M. Br{\"u}ggen]
       {T. A. En{\ss}lin$^1$, M. Br{\"u}ggen$^{1,2}$ \\
	$^1$ Max-Planck-Institut f{\"u}r Astrophysik, Garching D-85741,
	Germany\\
	$^2$ Institute of Astronomy, Madingley Road, Cambridge CB3 0HA, United Kingdom}
\date{Accepted ???? Month ??.
      Received  ???? Month ??;
      in original form  ???? Month ??}

\pagerange{\pageref{firstpage}--\pageref{lastpage}}
\pubyear{????}

\begin{document}

\maketitle

\label{firstpage}

\begin{abstract}
In several merging clusters of galaxies so-called {\it cluster radio
relics} have been observed. These are extended radio sources which do
not seem to be associated with any radio galaxies. Two competing
physical mechanisms to accelerate the radio emitting electrons have
been proposed: (i) diffusive shock acceleration and (ii) adiabatic
compression of fossil radio plasma by merger shock waves.  Here the
second scenario is investigated.  We present detailed 3-dimensional
magneto-hydrodynamical simulations of the passage of a radio plasma
cocoon filled with turbulent magnetic fields through a shock
wave. Taking into account synchrotron, inverse Compton and adiabatic
energy losses and gains we evolved the relativistic electron
population to produce synthetic polarisation radio maps. On
contact with the shock wave the radio cocoons are first compressed and
finally torn into filamentary structures, as is observed in
several cluster radio relics. In the synthetic radio maps the
electric polarisation vectors are mostly perpendicular to the
filamentary radio structures.  If the magnetic field inside the cocoon
is not too strong, the initially spherical radio cocoon is transformed
into a torus after the passage of the shock wave. Very recent,
high-resolution radio maps of cluster radio relics seem to exhibit
such toroidal geometries in some cases. This supports the
hypothesis that cluster radio relics are fossil radio cocoons that
have been revived by a shock wave. For a late-stage relic the ratio of
its global diameter to the filament diameter should correlate with the
shock strength.  Finally, we argue that the
total radio polarisation of radio relic should be well correlated
with the 3-dimensional orientation of the shock wave that produced
the relic.
\end{abstract}

\begin{keywords}
galaxies: clusters: general -- intergalactic medium -- shock waves --
magnetohydrodynamics -- radio continuum: general -- polarization
\end{keywords}

\section{Introduction}

\begin{figure}
\begin{center}
\psfig{figure=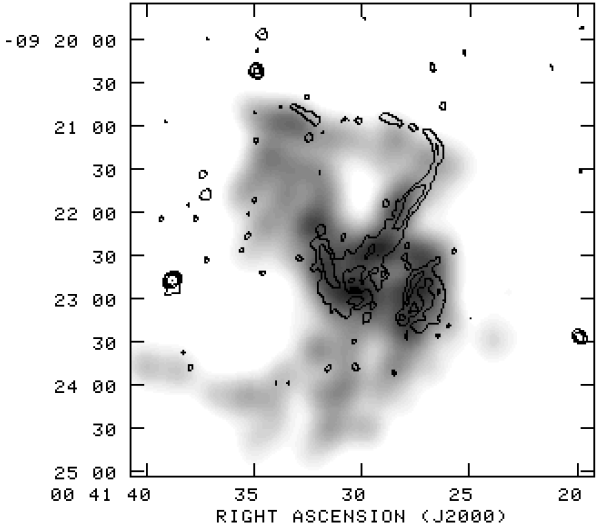,width=0.45 \textwidth,angle=0}
\end{center}
\caption[]{\label{fig:a85}Cluster radio relic in Abell 85 at 327 MHz
(grey-scale, Giovannini \& Feretti 2000\nocite{2000NewA....5..335G})
and at 1.4 GHz (contours, Slee et al. 2001\nocite{2001AJ....122.1172S}). The 327
MHz maps demonstrates that the thin 1.4 GHz filament extending
Northwards belongs in fact to a closed loop in projection. The two
smaller, southern blobs on the 1.4 GHz map have also torus-like
morphologies.}
\end{figure}

In the current picture of hierarchical structure formation clusters of
galaxies grow mainly by the merging of smaller and moderately sized
sub-clusters. During such merger events a significant fraction of the
kinetic energy of the inter galactic medium (IGM) is dissipated in
Mpc-sized shock waves. The shock waves are responsible for heating the
IGM to temperatures of several keV. Moreover, they are likely to inject
and accelerate relativistic particle populations on cluster scales.

Cluster-wide relativistic electron populations are indeed observed in
several merging or post-merging clusters. These electrons reveal their
presence by emitting synchrotron emission in the magnetic fields of
the IGM to form the so-called {\it cluster radio halos}. Cluster radio
halos have steep radio spectra (spectral index $\alpha \approx
1-1.5$); they are unpolarised, and their diffuse morphologies are
often similar to those of the thermal X-ray emission of the cluster
gas (Govoni et al.  2001)\nocite{govoni2001}.  A comparison of the
typical radiative lifetime of the radio emitting electrons, which is
of the order of 0.1 Gyr, to the typical dynamical timescale of the
cluster merger, which is of the order of 1 Gyr, shows that the
electron population of a halo cannot be produced directly by a merger
shock. Either these electrons have been re-accelerated more recently,
or they have been injected by a long-lived shock-accelerated proton
population via the hadronic charged pion production. For reviews on
recent observations of diffuse cluster radio sources see Feretti
(1999)\nocite{Feretti.Pune99}, Giovannini, Tordi, Feretti
(1999)\nocite{1999NewA....4..141G}, and Kempner \& Sarazin (2001)
\nocite{2001ApJ...548..639K}.

Unlike the radio halos, the so-called {\it cluster radio
relics}\footnote{The name {\it `cluster radio relic'} was originally
chosen in order to reflect the similarities of these sources with {\it
`relic radio galaxies'}. The latter are dying radio cocoons that are
no longer supplied with fresh radio plasma from the central parent
galaxy. In cluster relics, usually no parent galaxy can be
identified. Moreover, they have only been found in clusters of
galaxies, and therefore they were named {\it `cluster radio
relics'}. Recently, the connection between cluster radio relics and
radio galaxies has been debated. Alternative, possibly less misleading
names have been proposed, e.g.  {\it `radio flotsam'} by
R. Ekers. Here, we use the original terminology a) in order to be
consistent with the former literature and b) since our work provides
support for a scenario, in which these objects consist of
re-illuminated fossil radio plasma, so that they are indeed shock
processed relic radio galaxies.} are more directly related to cluster
mergers (En{\ss}lin et al. 1998, Roettiger et al. 1999, Venturi et al.
1999, En{\ss}lin \& Gopal-Krishna
2001)\nocite{1998AA...332..395E,1999ApJ...518..603R,1999dtrp.conf..27V,2001A&A...366...26E}. They
are also extended radio sources with a steep spectrum. In the
literature radio relics are often confused with radio halos even
though several distinctive properties exist. Cluster radio relics are
typically located near the periphery of the cluster; they often
exhibit sharp emission edges and many of them show strong radio
polarisation.

In several cases it could be shown that shock waves are present at the
locations of the relics. In Abell 2256 and Abell 1367
temperature substructures of the hot IGM could be detected (Briel \&
Henry 1994, Donnelly et
al. 1998)\nocite{1994Natur.372..439B,1998ApJ...500..138D} which points
towards the presence of shock waves at the location of the cluster
relics in these clusters. For Abell 754 (Roettiger at al. 1998; Kassim
et al. 2001)\nocite{1998ApJ...493...62R,kassim2001a}, Abell 2256
(Roettiger et al. 1995)\nocite{1995ApJ...453..634R}, Abell 3667
(Roettiger et al. 1999)\nocite{1999ApJ...518..603R} and also the Coma
cluster (Burns et al. 1994)\nocite{1994ApJ...427L..87B} numerical
simulations of mergers were fitted to the X-ray
data. These simulations predict shock waves at locations of
observed cluster radio relics.

The cluster radio relic 1253+275 in the Coma cluster shows a
morphological connection to the nearby radio galaxy NGC 4789
(Giovannini, Feretti \& Stanghellini
1991\nocite{1991A&A...252..528G}). This suggests that radio
relics may be fossil radio plasma that has been revived by a
shock. Fossil radio plasma is the former outflow of a radio galaxy in
which the high-energy radio emitting electrons have lost their
energy. Due to their invisibility in the radio these
cocoons are also called {\it radio ghosts} (En{\ss}lin
1999)\nocite{1999dtrp.conf..275E}.

The first relic formation models considered diffusive shock
acceleration (Fermi I) as the process producing the radio emitting
electrons (En{\ss}lin et al. 1998, Roettiger et al. 1999, Venturi et
al.
1999)\nocite{1998AA...332..395E,1999ApJ...518..603R,1999dtrp.conf..27V}.
However, when a fossil radio cocoon is passed by a cluster merger
shock wave, with a typical velocity of a few 1000 km/s, the cocoon is
compressed adiabatically and not shocked because of the much
higher sound speed within it. Therefore, shock acceleration cannot be
the mechanism that re-energises the relativistic electron
population. But the energy gained during the adiabatic compression
combined with the increase in the magnetic fields strength can
cause the fossil radio cocoon to emit radio waves again. One
prerequisite for this is that the electron population is not older
than 0.2 - 2 Gyr (En{\ss}lin \& Gopal-Krishna
2001\nocite{2001A&A...366...26E}). The timescale depends on the
conditions in the surroundings, mainly the external pressure. In
high-pressure environments, such as in cluster cores, the synchrotron
losses are expected to be much higher due to the higher
internal magnetic fields of pressure-confined radio
plasma. This leads to a shorter maximum age for the fossil radio
plasma if it is to be revived. Therefore, relics are found
preferentially at the periphery of clusters where the pressure is
lower. Moreover, numerical simulations show that merger shocks are
found more frequently in peripheral cluster regions than in the denser
cores (Quilis, Ibanez, \& Saez 1998\nocite{1998ApJ...502..518Q};
Miniati et al. 2000\nocite{2000ApJ...542..608M}). Both effects, the
longer time for radio plasma to be revivable, and the higher frequency
of shock waves could explain why cluster radio relics are more
frequently observed at the outskirts of clusters than in more central
regions.

En{\ss}lin \& Gopal-Krishna (2001)\nocite{2001A&A...366...26E} showed
that the spectral properties of cluster radio relics are well
reproduced by this scenario. Here, we demonstrate that the observed
morphologies and polarisation patterns are reproduced by this model as
well. This is done with the help of the first 3-dimensional
magneto-hydrodynamical (MHD) simulations of a fossil radio cocoon that
is passed by a shock wave. We produce artifical radio maps that can be
compared directly to high-resolution radio maps of relics.

\begin{figure*}
\psfig{figure=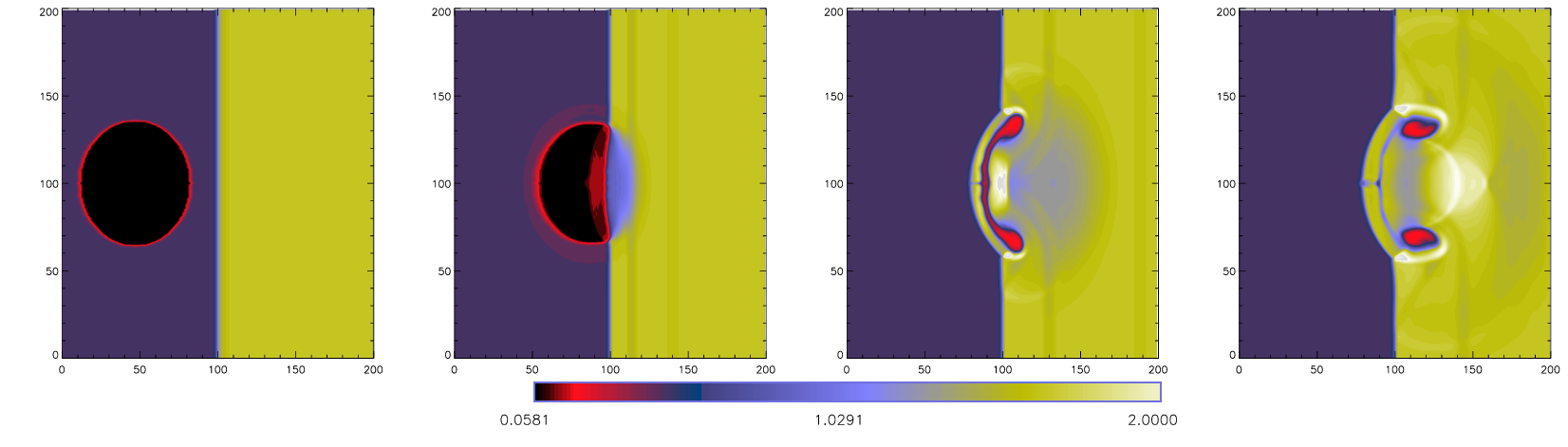,width=1.0 \textwidth,angle=0}
\psfig{figure=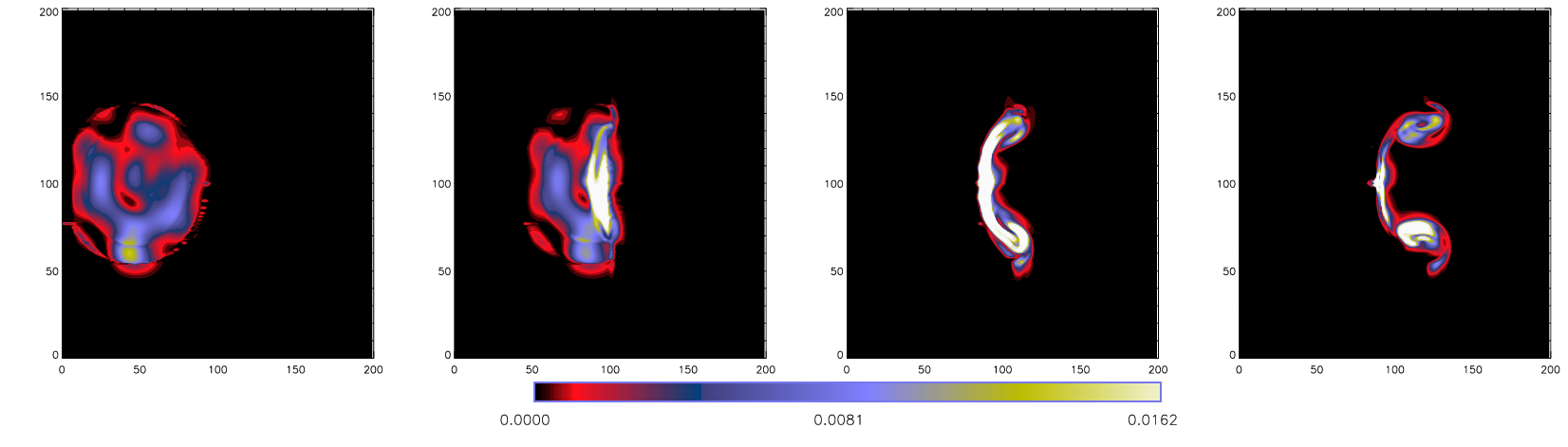,width=1.0 \textwidth,angle=0}
\caption[]{\label{fig:evolution1} Evolution (from the initial
configuration (left) to the final one (right)) of the gas density
(top) and magnetic field energy density (bottom) in a central slice
through the simulation volume. The pre-shock region is on the left
hand side, and the post-shock region on the right hand side of the
stationary shock wave, which is located at the centre of the
computational box. The formation of a torus is best visible in the
density plot. In this simulation the projection of the vector
potential of the magnetic fields is not done by Eq. \ref{eq:proj}, but
by the modified form ${\mathbf A} = [{\mathbf I}-g(r)({\mathbf
I}-\hat{\mathbf r}\hat{\mathbf r})]/(1-2\,g(r)/3)\cdot\tilde{\mathbf
A}$ in order to have a more uniform magnetic field within the radio
cocoon. The numbers on the axes label the zones of the grid.}
\end{figure*}

\section{Method}
\subsection{3D MHD Simulation}

The magneto-hydrodynamical simulations were obtained using the ZEUS-3D
code which was developed especially for problems in astrophysical
hydrodynamics (Stone \& Norman 1992a,
b)\nocite{1992ApJS...80..753S,1992ApJS...80..791S}.  The code uses
finite differencing on a Eulerian grid and is fully explicit in
time. It is based on an operator-split scheme with piecewise linear
functions for the fundamental variables. The fluid is advected through
a mesh using the upwind, monotonic interpolation scheme of van Leer.
The magnetic field is evolved using a modified constrained transport
technique which ensures that the field remains divergence-free to
machine precision. The electromotive forces are computed via upwind
differencing along Alfv\'en characteristics. For a detailed
description of the algorithms and their numerical implementation see
Stone \& Norman (1992a, b).

The simulations were computed on a Cartesian grid with 100$^3$ and
200$^3$ equally spaced zones. The fluid flows in the $x$-direction
with inflow boundary conditions at the lower boundary and outflow
conditions at the outer boundary. The boundary conditions in the $y$-
and $z$-directions were chosen to be reflecting.  The simulation was set up such that a stationary shock
formed at the centre of the simulation box that is perpendicular
to the direction of the flow.

In the pre-shock region a spherical bubble of radius $R$ was set up,
in which the density was lowered by a factor of 10 with respect to the
environment. In turn, the temperature in the bubble was raised such
that the bubble remained in pressure equilibrium with its
surroundings. Inside the bubble a magnetic field was set up as
follows: The magnetic vector potential $\tilde{\mathbf A}$ was
composed of 125 Fourier modes with random phases and
amplitudes. $\tilde{\mathbf A}$ was then scaled as $k^{-3}$, where $k$
is the wavenumber of the mode. This leads to a magnetic power
spectrum of $P_B(k)\, dk\sim k^{-1}\,dk$.  To avoid leakage of
the magnetic field into the surroundings of the bubble, we employed
the projection:

\begin{equation}
\label{eq:proj}
{\mathbf A} = [{\mathbf I}-g(r)({\mathbf I}-\hat{\mathbf
r}\hat{\mathbf r})]\cdot\tilde{\mathbf A} ,
\end{equation}
where ${\mathbf I}$ is the identity, $r$ the radius measured from the
centre of the bubble, $\hat{\mathbf r}$ the corresponding unit vector and
$g(r)=(r/R)^2$ if $r<R$ and $g(r)=1$ if $r\geq R$.

The bubble was filled with around $10^4$ uniformly distributed tracer
particles that are advected with the flow.  We employed a polytropic
equation of state with an adiabatic index of $\gamma=5/3$. The
simulations with the serial version of the code were performed on a
SUN ULTRA 10 workstation and the parallel version of the code was run
on 8 processors on an SGI ORIGIN 3000.

Finally, we should address some issues related to the accuracy of
these kinds of finite-difference hydrodynamical simulations.  First,
one can note that the gas density in the bubble increases with time. Any
observed diffusion is entirely numerical. The boundary between the
bubble and the ambient medium also becomes less sharp as the
simulation proceeds due to discretisation errors in the advection
scheme. For a test of the advection algorithm in the ZEUS code see
Stone \& Norman (1992a,b). Simple tests showed that during the
advection of a sharp discontinuity over a grid of 200 zones the
discontinuity is spread over 3-4 grid cells. Therefore, the smaller
features in our simulation are more susceptible to advection errors
than the larger ones.

Second, numerical viscosity is responsible for suppressing small-scale
instabilities at the interface between the bubble and the cooler,
surrounding gas.  To assess the effects of numerical viscosity, we
have repeated our simulations in 2D on grids with 300$^2$ and 500$^2$
grid points. Finally, we found that numerical resistivity destroys the
magnetic field upon compression by the shock. With the affordable
numerical resolution the magnetic field orientation reversals are
pressed in very nearby computational cells. This is discussed further
in Sec. 3.2.

\begin{figure}
\begin{center}
\psfig{figure=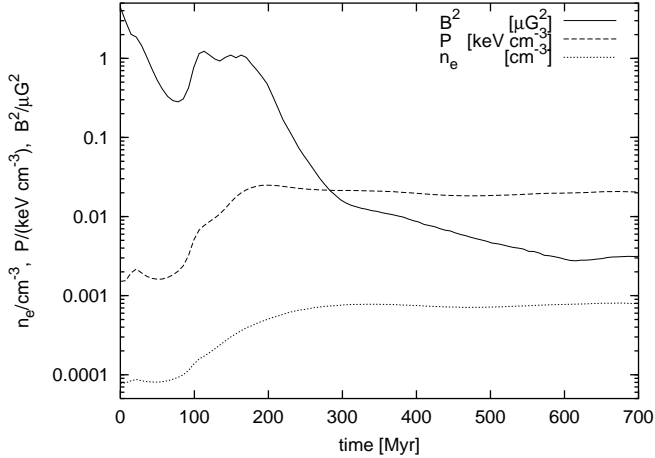,width=0.5 \textwidth,angle=0}
\end{center}
\caption[]{\label{fig:evolution.g53r33Bm} Evolution of the average
magnetic field energy density $\sim B^2$, the thermal pressure
$P$, and the electron number density $n_{\rm e}$ at the locations of
the tracer particles in the {\it sSwB} simulation (sSwB = strong
Shock, weak B-field; see Sect. \protect{\ref{sec:results}}) for
notation . The impact of the shock compression is visible after
around 100 Myr. Also the rapid decline of the magnetic field strength
due to numerical resistivity is visible. This decline is not observed
in 2-d test simulations with either a much higher resolution or a
much more ordered magnetic field.}
\end{figure}

\subsection{Radio Maps}

The radio maps are constructed using around $10^4$ passively advected
tracer particles. Initially, each tracer particle is located inside
the radio plasma cocoon and is associated with the same initial
relativistic electron population. Then the electron spectrum for each
tracer particle is evolved in time taking into account synchrotron,
inverse Compton, and adiabatic energy losses and gains. In the
formalism of En{\ss}lin \& Gopal-Krishna
(2001)\nocite{2001A&A...366...26E}\footnote{Equivalent formalisms were
derived and applied to numerical simulations by other authors,
e.g. Matthews \& Scheuer (1990a,b)
\nocite{1990MNRAS.242..616M,1990MNRAS.242..623M}, and Churazov et
al. (2001) \nocite{2001ApJ...554..261C}.} this spectral evolution requires
two quantities to be tracked as a function of time. The first is the
compression history, i.e.
\begin{equation}
{C}(t) = V_{0}/V(t) ,
\end{equation}
where $V_{0}$ and $V(t)$ are the initial and present specific volumes
of the tracer particles. In our calculations we use the pressure
$P(t)$ at the location of the tracer particles to work out the
compression via the equation of state:
\begin{equation}
\label{eq:C(P)}
{C}(t) = (P(t)/P_{0})^{1/\gamma_{\rm rp}} ,
\end{equation}
where $\gamma_{\rm rp}$ is the adiabatic index of radio
plasma. Even though our MHD simulations have a constant adiabatic index of
$5/3$, we use $\gamma_{\rm rp} = 4/3$ in Eq. \ref{eq:C(P)} in order to
simulate the higher energy gains of the electrons in the
more compressible relativistic radio plasma.

\begin{figure}
\begin{center}
\psfig{figure=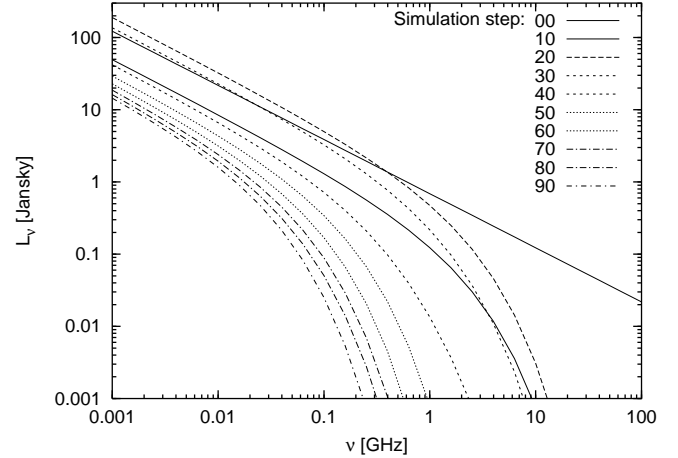,width=0.5 \textwidth,angle=0}
\end{center}
\caption[]{\label{fig:spec43_g53_r33_Bm} Evolution of the radio
spectrum of the {\it sSwB} simulation. The numbers (00, 10, ...,
90) label different simulation data-dump time-steps. A single time
step is approximately 7 Myr long. Clearly visible is the decrease of
the cutoff frequency with time. Only in the interval 10-20 (70-140
Myr) the shock compression is able to reverse this. During this phase
also the overall emissivity at lower radio frequencies increases
mainly due to field strength amplification. At all other time-steps
the low frequency flux decreases artificially, due to magnetic field
annihilation by numerical field diffusion.}
\end{figure}

The second quantity that has to be tracked in order to evolve the
electron spectrum is the energy (or momentum) of an electron that was
initially injected with infinite energy. This cutoff momentum $p_*$ of
the electron spectrum (here and in the following dimensionless units
for electron momenta $p = P_{\e} /(m_\e c)$ are used, where $m_\e$ is
the electron rest-mass, and $c$ the speed of light) is given by
\begin{equation}
\label{eq:mysol}
\frac{1}{p_*(t)} = a_{0} \, \int_{t_{0}}^t \!\!\!\! dt'
\,(u_B(t')+u_C(t'))\, \left( \frac{{C}(t')}{{C}(t)}
\right)^{\frac{1}{3}}\,.  
\end{equation}
Here $a_{0} = \frac{4}{3}\, \sigma_{\rm T}/(m_\e\, c)$ with
$\sigma_{\rm T}$ the Thomson cross-section, and $u_B$ and
$u_C$ are the magnetic and cosmic microwave background energy
densities, respectively.  Hence the energy evolution of any electron
can be calculated using
\begin{equation}
p(p_0, t) = \frac{p_{0}}{{C}(t)^{-\frac{1}{3}} + p_{0} / p_*(t)}\,.
\end{equation}

The electron spectrum per dimensionless momentum of a tracer particle
$f(p,t)\, dp$ for $p<p_*(t)$ is given by
\begin{equation}
\label{eq:spec1}
f(p,t) = f_{0}(p_{0}(p,t)) \frac{\partial p_{0}(p,t)}{\partial
p}\,,
\end{equation}
where
\begin{equation}
p_{0}(p,t) = p\,{C}(t)^{-\frac{1}{3}}/(1 -
p/p_*(t))\,.
\end{equation}
If the original distribution function is a power-law
\begin{equation}
f_{0}(p_{0}) = {g}_{0} \, p_0^{-\alpha_\e}
\end{equation}
for $p_{\rm min\,0} < p_0 < p_{\rm max\, 0}$ the resulting spectrum is
\begin{equation}
\label{eq:spec2}
f(p,t) = {g}_{0} \, {C}(t)^{\frac{\alpha_\e -1}{3}}\,
p^{-\alpha_\e}\, \left( 1 - p/p_{*}(t) \right)^{\alpha_\e -2}
\end{equation}
for $p_{\rm min}(t)= p(p_{\rm min\,0},t) < p < p_{\rm max}(t) =
p(p_{\rm max\, 0},t)$.

At fixed intervals during the simulation (up to 100) the positions of
the tracer particles as well as the gas and magnetic field properties
were written out. These were then used to tabulate $C(t)$ and $p_*(t)$
interpolating linearly in the integration in Eq. \ref{eq:mysol}. A
second programme read in these tables to calculate the Stokes I-, Q-,
and U-polarisation radio maps using the standard integration kernels
of synchrotron radiation theory (Rybicki \& Lightman
1979\nocite{1979rpa..book.....R}). At this step we have specified the
parameters of the initial electron spectrum, viewing angle, distance
and observing frequency. The maps were stored in the FITS format, and
visualised with the radio-astronomical data reduction package AIPS.

\section{Results\label{sec:results}}
\label{sec:results}

In the following, results from four simulation runs are presented: two
simulations with a weak shock (external compression factor
$C_{\rm gas} = 2$), and two with a stronger shock ($C_{\rm
gas} = 3.3$). For each shock strength, simulations with dynamically
unimportant (`weak') and dynamically important (`strong') field strengths
are discussed. In those simulations with strong magnetic fields, the
magnetic pressure initially exceeds the thermal pre-shock
pressure. But a short violent expansion quickly establishes
approximate pressure equilibrium. In the following, we name the
simulation with the weak shock and the weak field strength `{\it
wSwB}', with the weak shock and the strong field strength '{\it
wSsB}', and likewise.

Physical units are assumed such that the simulated radio relic
resembles one which is located in the denser regions of a small galaxy
cluster. In principle, we could assume physical units that correspond
to more peripheral relics that reside in lower pressure
environments. But in this case the artificially decaying magnetic
fields (see Sec. \ref{sec:global}) have a greater effect on the radio
luminosity because the inverse Compton losses are bigger than the
synchrotron losses.

The pre-shock external gas density is set to about $5\cdot
10^{-4}\,{\rm electrons/cm}^3$ and the temperature to 1-2 keV. The
simulation box is assumed to have a size of 1 Mpc$^3$ and to be
located at a distance of 100 Mpc from the observer.

\begin{figure}
\begin{center}
\psfig{figure=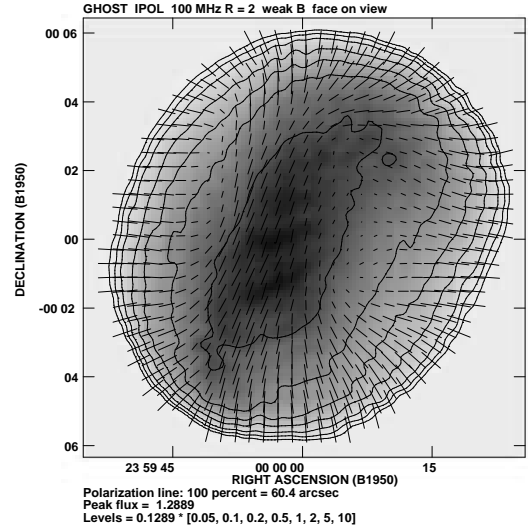,width=0.40 \textwidth,angle=0}
\end{center}
\caption[]{\label{fig:map.WW+10-00} Radio emission of the radio bubble
in the model {\it wSwB} before shock passage. The polarisation
E-vectors are displayed by dashes with the length proportional to the
relative polarisation. Here and in the following radio maps the flux
is given in Jansky per simulation pixel (with linear size of 20.6
arcsec)}
\end{figure}

\subsection{Torus Formation}

The passage of a radio cocoon through a shock wave can be seen in
Fig. \ref{fig:evolution1}. The formation of a torus can be explained
as follows: At a shock wave, the ram-pressure of the
pre-shock gas (plus its small thermal pressure) and the thermal
pressure of the post-shock gas are in balance. But when the lighter
radio plasma comes into contact with the shock surface, the ram
pressure is reduced at the point of contact. The post-shock gas starts
to expand into the volume occupied by the still low pressure radio
plasma. This can be seen in the second and third displayed panel
of Fig. \ref{fig:evolution1}. Finally, the radio plasma is disrupted
and a torus forms.

A possible evolutionary scenario of the radio morphology of the cocoon
could be similar to what is shown in Fig. \ref{fig:evolution1}. In
the early stages of the compression the torus has not formed and one
expects a sheet like radio relic to be visible. The relic should get
more and more edge brightened while the radio plasma accumulates in
the torus. Finally, when the thin sheet is disrupted, only a
torus-like object should be visible.

Thus we expect that the early, sheet-like radio relics have
a `younger', less steepened spectrum, while the later-stage,
filamentary or toroidal relics have a much higher spectral age with
a bent, steep radio spectrum.

These predictions are supported by the fact that the relatively
sheet-like relics in the cluster Abell 2256 have a flat spectrum
(R{\"o}ttgering et al. 1994\nocite{1994ApJ...436..654R}) whereas all
the filamentary relics in the sample observed by Slee et
al. (2001)\nocite{2001AJ....122.1172S} exhibit steep, bent and
therefore `aged' spectra. We note that this statement needs additional
observational confirmation.

\begin{figure}
\begin{center}
\psfig{figure=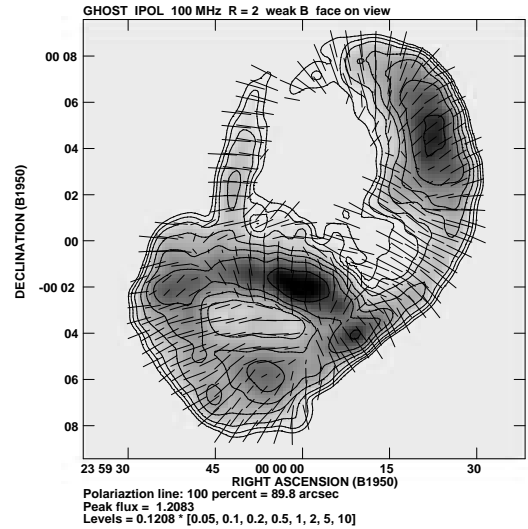,width=0.40 \textwidth,angle=0}
\end{center}
\caption[]{\label{fig:map.WW+70-00} Same as Fig. \ref{fig:map.WW+10-00}. Late
stage of the shock passage in the same model {\it wSwB}.}
\end{figure}

\begin{figure}
\begin{center}
\psfig{figure=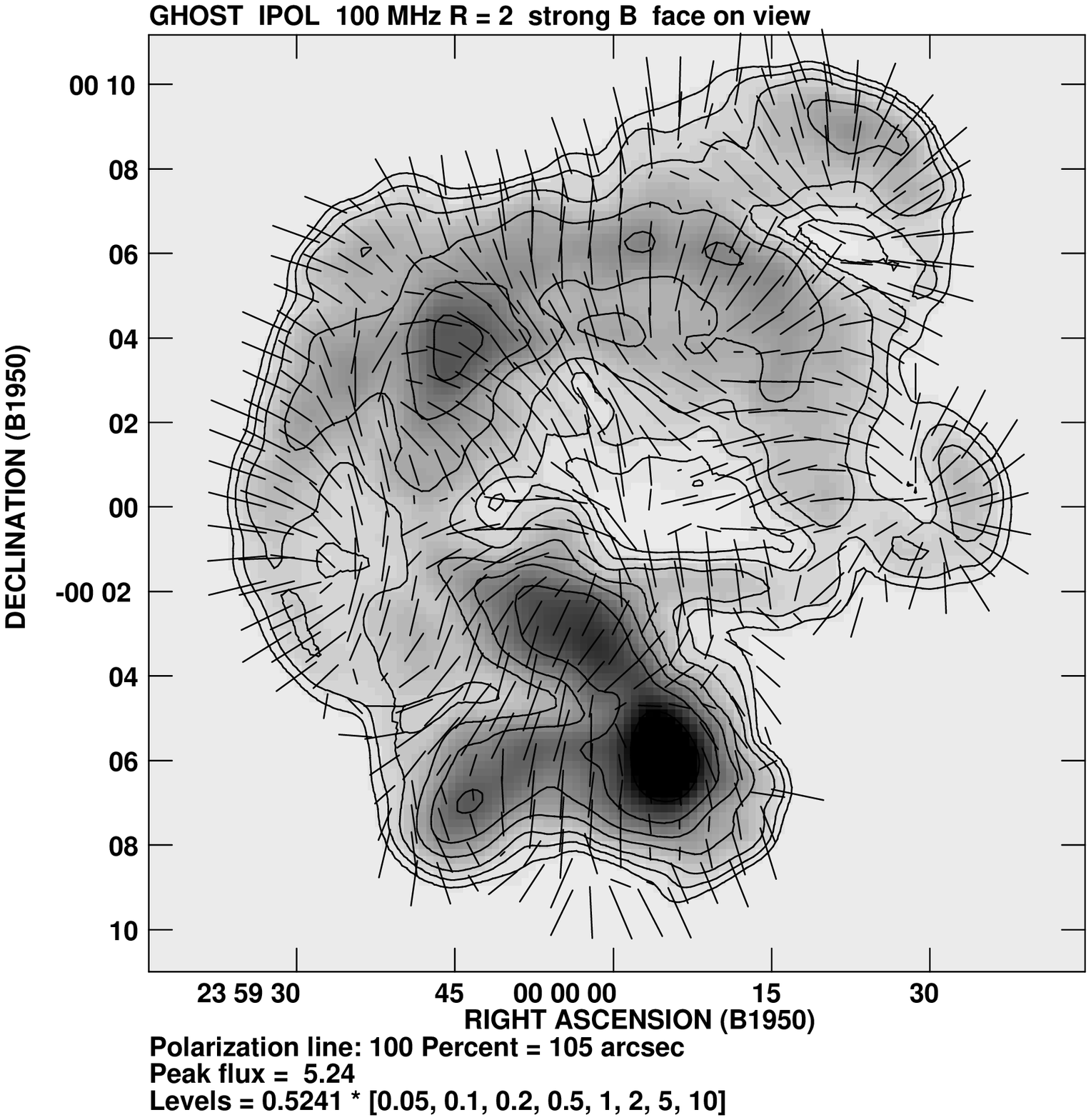,width=0.40 \textwidth,angle=0}
\end{center}
\caption[]{\label{fig:map.WS60-00} Face-on view of the late stage of
the shock passage in the model {\it wSsB}. For details see
Fig. \ref{fig:map.WW+10-00}.}
\end{figure}

\begin{figure}
\begin{center}
\psfig{figure=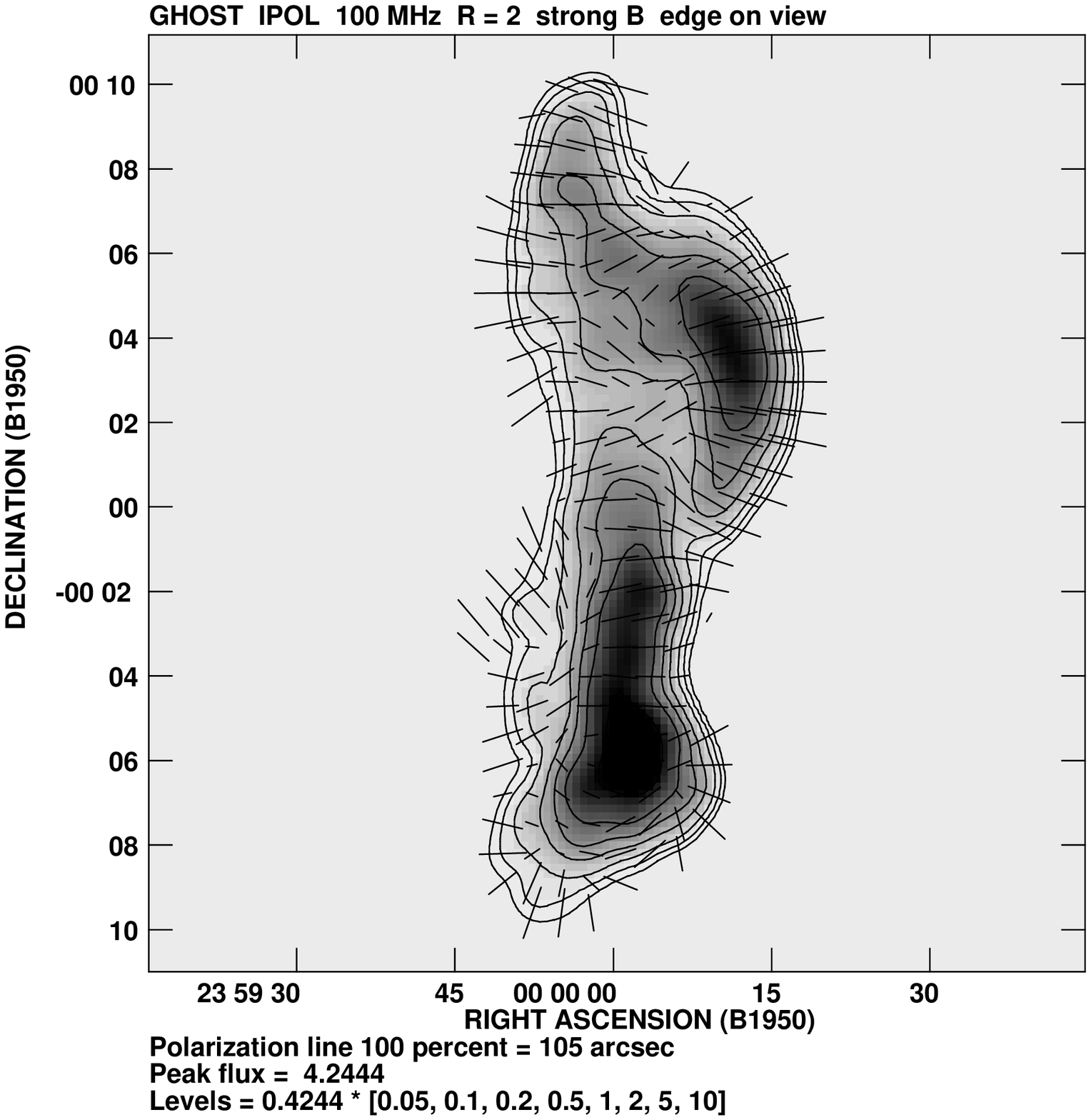,width=0.40 \textwidth,angle=0}
\end{center}
\caption[]{\label{fig:map.WS60-90} Edge-on view of the late stage of
the shock passage in the model {\it wSsB}. For details see
Fig. \ref{fig:map.WW+10-00}.}
\end{figure}

\begin{figure}
\begin{center}
\psfig{figure=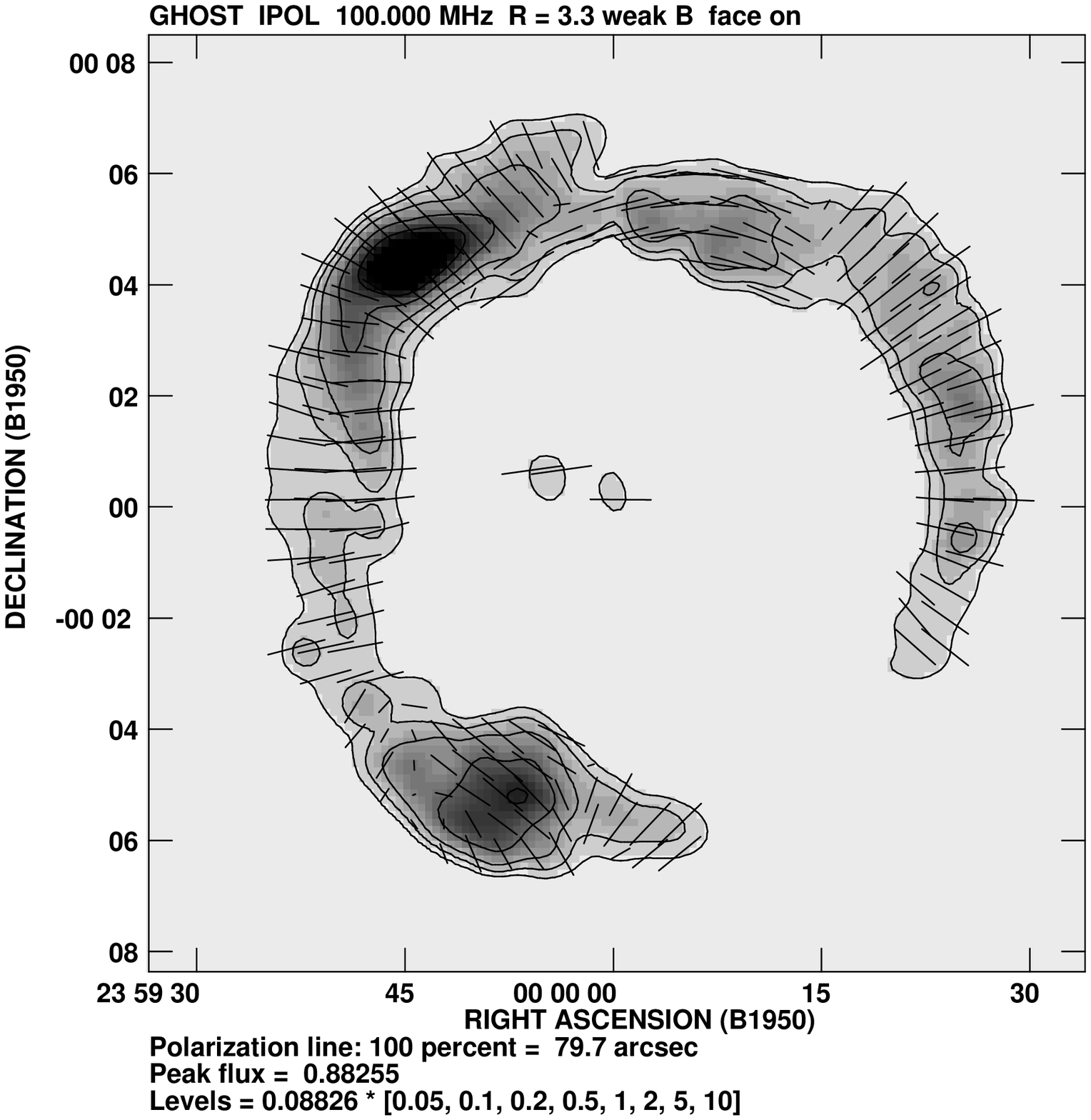,width=0.40 \textwidth,angle=0}
\end{center}
\caption[]{\label{fig:map.SW70-00} Face-on view of the late stage of
the shock passage in the model {\it sSwB}. For details see
Fig. \ref{fig:map.WW+10-00}.}
\end{figure}

\begin{figure}
\begin{center}
\psfig{figure=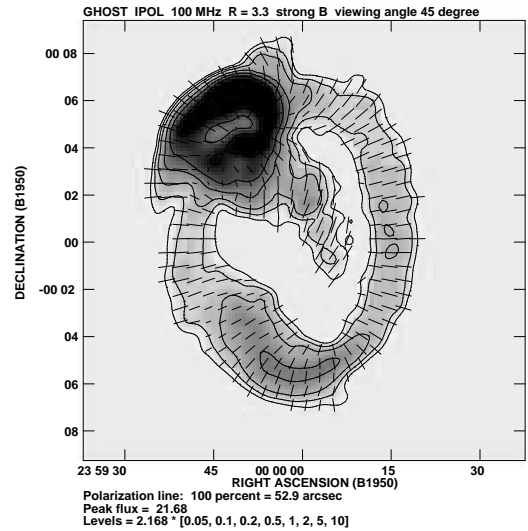,width=0.40 \textwidth,angle=0}
\end{center}
\caption[]{\label{fig:map.SS+70-45} 45 degree view of the late stage of
the shock passage in the model {\it sSsB}. For details see
Fig. \ref{fig:map.WW+10-00}.}
\end{figure}

\subsection{Global Properties \label{sec:global}}

In Fig. \ref{fig:evolution.g53r33Bm} the evolution of the average
magnetic field strength, the thermal pressure and the electron number
density at the location of the tracer particles are shown for one
simulation. The bubble reaches its maximal compression after
75-200 Myr. The associated increase in the thermal electron
density is somewhat higher than predicted by simple adiabatic
compression. This is probably due to gas electrons that have
diffused numerically into the radio plasma volume.  However,
this does not affect our radio maps. The spatial distribution of the
relativistic particles is solely determined by the locations of tracer
particles. The compressional state of the relativistic electrons is
calculated from the fluid pressure and not from the density in order
to be less sensitive to numerical diffusion.

Moreover, the decline of the magnetic field strength is caused by the
numerical resistivity (or magnetic diffusivity) of the code. The
reason for the strong impact of the numerical diffusion of the
magnetic field on our results lies in the thin morphology the radio
plasma after the shock passage. This brings low density and high
density regions, and also regions with differently oriented magnetic
fields close together, producing strong gradients which are sensitive
to numerical diffusion. Since the radio plasma is compressed into
filaments that are only a few cells wide, magnetic diffusion leads to
strong artificial annilihation (or reconnection) of magnetic fields.

Two-dimensional resolution studies indicate that this artificial
decrease of the magnetic field strength should become small for a
$500^3$ cell simulation. Since such a simulation is at present not
feasible for us we will try to focus on robust conclusions that are
not severely affected by this inaccuracy. However, we have made
experiments with low resolution and large-scale fields in two
dimensions. These experiments revealed that the magnetic field
strength increases drastically upon compression and, subsequently,
remains strong -- as is expected in the absence of numerical
diffusivity.

\begin{figure}
\begin{center}
\psfig{figure=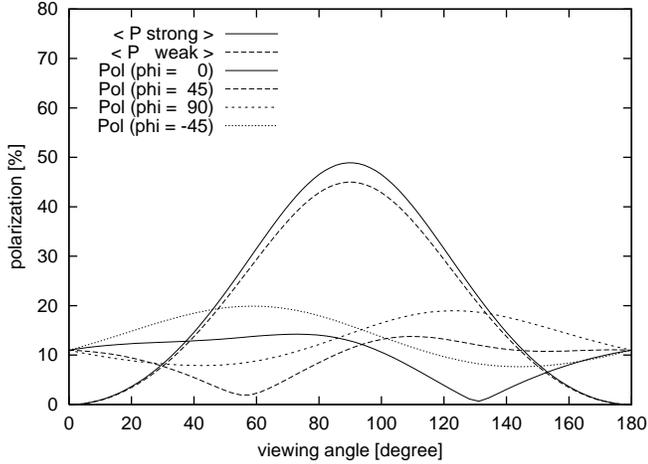,width=0.5 \textwidth,angle=0}
\end{center}
\caption[]{\label{fig:pol.WS20.100MHz} 100 MHz polarisation scans
around the radio source in the simulation {\it wSsB} at an early stage
of the shock passage with the viewing angle ranging from face-on
(viewing angle = $0^\circ$) over edge-on (viewing angle = $90^\circ$)
to viewing the back (viewing angle = $180^\circ$). The different
angles $\phi = -45, 0, +45, +90$ label four different routes of the
scans around the relic. }
\end{figure}

\begin{figure}
\begin{center}
\psfig{figure=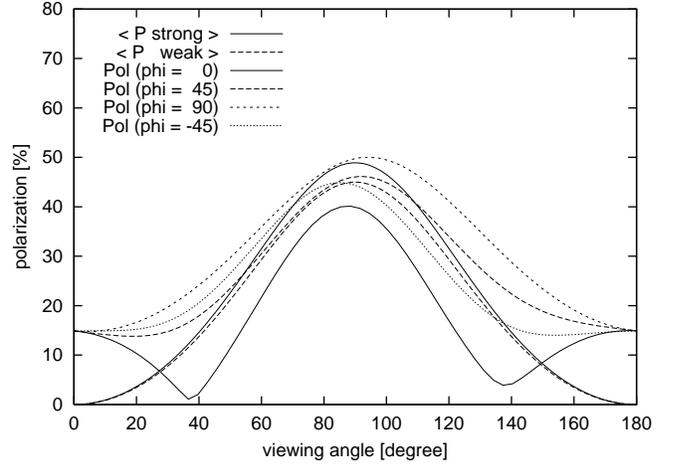,width=0.5 \textwidth,angle=0}
\end{center}
\caption[]{\label{fig:pol.WS60.100MHz} Same as Fig. \ref{fig:pol.WS20.100MHz},
but for a late stage of the shock passage.}
\end{figure}

\begin{figure}
\begin{center}
\psfig{figure=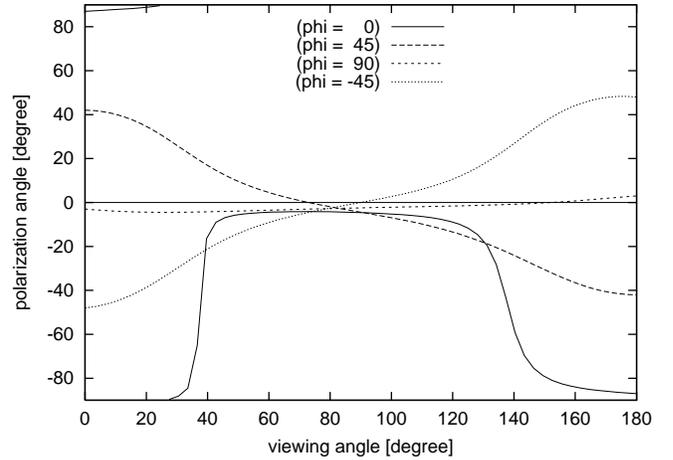,width=0.5 \textwidth,angle=0}
\end{center}
\caption[]{\label{fig:pol.ang.WS60.100MHz} Same as Fig.
\ref{fig:pol.WS60.100MHz} but here the angle between the projected
normal of the shock wave and the E-polarisation vector is displayed. Values
near zero mean that the projected magnetic fields are mostly aligned
with the projected shock orientation. }
\end{figure}

\subsection{Spectral Aging}

An increase of a factor $\approx 5$ in the radio luminosity during
the compression between 70 to 140 Myr can be seen in
Fig. \ref{fig:spec43_g53_r33_Bm}.  However, the greater
increase expected from the analytical model of En{\ss}lin \&
Gopal-Krishna (2001)\nocite{2001A&A...366...26E} could not be
reproduced in our simulations. This is a result of the decaying
magnetic fields due to numerical resistivity, and not a failure of the
model. In order not to be too much affected by the rapidly evolving
spectral cutoff we restrict our analysis in the following to a low
radio frequency, namely 100 MHz.

\subsection{Radio Morphology}

As the shock wave passes the initially spherical radio cocoon (Fig.
\ref{fig:map.WW+10-00}), the cocoon is torn into a filamentary
structure (Figs.  \ref{fig:map.WW+70-00}, \ref{fig:map.WS60-00},
\ref{fig:map.WS60-90}, \ref{fig:map.SW70-00}, and
\ref{fig:map.SS+70-45}). In the simulations with weak magnetic fields
the final morphology is toroidal for the strong shock (simulation {\it
sSwB}, Fig. \ref{fig:map.SW70-00}) and shows two tori in the case of a
weak shock wave ({\it wSwB}, Fig. \ref{fig:map.WW+70-00}). Such a
double torus seems to be observed in the relic in Abell 85 (Slee et
al. 2001\nocite{2001AJ....122.1172S}, and Fig. \ref{fig:a85}). The
degree of polarisation is relatively high everywhere. The electric
polarisation vectors tend to be perpendicular to the radio filaments
indicating aligned magnetic field structures within them. The
reason for this is that the compression is mostly perpendicular to the
filaments since this is how they formed. Therefore, the parallel field
component is preferentially enhanced.

But in some few spots (see Fig. \ref{fig:map.SW70-00}) the
polarisation E-vectors are aligned with the radio filaments. We
presume that these spots coincide with regions where the initial
magnetic component parallel to the shock was too weak to be
amplified above the perpendicular components.  Such
reversals are indeed observed in polarisation maps of relics but
multi-frequency observations still have to confirm that they are
intrinsic, and not induced by foreground Faraday rotation (R. Slee,
private communications).

In those simulations where the magnetic fields are dynamically
important nearly always a more complicated morphology is produced. In
the weak shock simulation ({\it wSsB}, Figs.  \ref{fig:map.WS60-00}
and \ref{fig:map.WS60-90}) the cocoon does not show any toroidal
structure after shock passage. But filamentary structures can still be
identified and the E-polarisation vectors seem to be approximately
perpendicular to the filaments. The edge-on projection shows that the
radio emitting volume is not restricted to the shock plane but also
shows kinks and extentions in the flow direction. Similar looking
kinks are observed in the radio relics in the cluster Abell
3667. These relics are also likely to be seen edge-on as the merger
shock lies within or close to the plane of the sky (Roettiger
1999)\nocite{1999ApJ...518..603R}.

In the simulation with a strong shock and strong fields ({\it sSsB}) a
large torus still appears, but a second smaller one is dominating the
total radio emission (Fig. \ref{fig:map.SS+70-45}). The extreme
brightness of this feature is likely to be an artifact of an unlucky
realization of the initial random magnetic field configuration than a
typical feature of a real radio relic, since a corresponding bright
spot already appeared in the initial radio cocoon.

\subsection{Radio Polarisation}

Our simulations indicate that for a sufficiently turbulent initial
magnetic field geometry the total integrated polarisation 
of a radio relic observed face-on should be small. It is of the
order of the total initial fractional polarisation of the source.  In
the case of a simulated toroidal relic all polarisation
orientations are roughly equally present and therefore cancel each
other out in the surface integration. But even in a more complex
relic, as shown Fig. \ref{fig:map.WS60-00}, the integrated
polarisation cancels out in the face-on view.

In the edge-on view this is different. Now a preferential direction
exists: E-vectors tend to be aligned with the projected shock normal.

It is interesting to study how reliably the overall polarisation
strength and direction can be used as a measure of the 3-dimensional
shock orientation. 

Analytic predictions for the polarisation strength of shock compressed
random magnetic fields were already derived in Laing (1980) and
En{\ss}lin et
al. (1998)\nocite{1980MNRAS.193..439L,1998AA...332..395E}. In these
derivations it was assumed that the compression is linear along the
flow axis and that the fields follow completely passively. The
resulting percentage radio polarisation is then given by:
\begin{equation}
\label{eq:Pweak}
<P_{\rm weak}> = \frac{s +1}{s +\frac{7}{3}}
\,\,\, \frac{\sin^2\delta }{\frac{2\, C_{\rm gas}^2}{C_{\rm gas}^2-1}-\sin^2\delta}\,\,.
\end{equation}
The term $C_{\rm gas}$ denotes the shock compression factor, $s$ the
spectral index of the electron population, and $\delta$ the viewing
angle.  If the magnetic fields dominate the total pressure one can
allow them to expand to pressure equilibrium after shock passage. In
this case the polarisation
\begin{equation}
\label{eq:Pstrong}
<P_{\rm strong}> = \frac{s +1}{s +\frac{7}{3}}
\,\,\, \frac{\sin^2\delta }{\frac{2}{15}\,\,\frac{13 C_{\rm gas}-
7}{C_{\rm gas}-1}-\sin^2\delta}
\end{equation}
is surprisingly similar to that expected in the weak field case
(En{\ss}lin et al. 1998\nocite{1998AA...332..395E}). This can be seen
from the reference curves in Fig. \ref{fig:pol.WS20.100MHz} and
\ref{fig:pol.WS60.100MHz}.

Also shown in these figures are viewing angle scans of the total
polarisation of the radio emission in the simulation {\it wSsB} before
(Fig. \ref{fig:pol.WS20.100MHz}) and after (Fig.
\ref{fig:pol.WS60.100MHz}) shock passage. These figures show that, in
addition to the intrinsic polarisation pattern of the turbulent
fields, the shock passage imprints a strong polarisation signature
onto the radio plasma.

It is remarkable that the characteristics of the simulated radio
polarisation are similar to the analytic models of shocked magnetic
fields. If the amplitude of the polarisation curves in our simulations
are not too strongly affected by artifacts, such as the decaying field
strength, or the incorrect adiabatic index of the radio plasma, we can
conclude that the shock acceleration and the revived fossil radio
plasma are not distinguishable by this. Alternatively, the simplistic
analytic theory of shocked fields seems to be a good guide in relating
viewing angle and radio polarisation.

The reason for the similar polarisation pattern in the different
models is likely that the magnetic fields in all cases
are nearly completely aligned with the shock plane. In the extreme case of
completely shock plane aligned fields, but with randomly distributed
orientations within that plane, the polarisation is given by
\begin{equation}
\label{eq:Pmax}
<P_{\rm max}> = \frac{s +1}{s +\frac{7}{3}}
\,\,\, \frac{\sin^2\delta }{2-\sin^2\delta}\,.
\end{equation}
This is not too far from the other models.

Whenever the imprinted polarisation dominates over the initial
intrinsic one, the direction of the E-vector can be used relatively
reliably to infer the shock normal projected onto the plane of the sky
(Fig. \ref{fig:pol.ang.WS60.100MHz}). The angle with respect to this
plane can be estimated roughly from the total polarisation.

Thus, in principle, the 3-dimensional orientation (modulo a mirror
ambiguity) can be derived from the polarisation data only.  However,
these points will need additional investigations before it is
applicable to real data. Especially the role of the initial field
geometry has to be investigated. Our simulated radio ghosts showed
deviations from the analytic predictions on the order of the total
initial polarisation of the source. Such deviations may significantly
reduce the accuracy of any polarisation based viewing angle
estimate. However, we think that according to our model there
should be a correlation between total polarisation and the viewing
angle due to the field ordering action of the shock wave. See
En{\ss}lin et al. (1998)\nocite{1998AA...332..395E} for an early
attempt to compare these quantities for a small smaple of cluster
radio relics.

\subsection{Shock Properties}

On the basis of our simulations we speculate that the observed
dimensions of a cluster radio relic with a toroidal shape may be
used to get a rough estimate of the shock strength. This is based on
the observation that in the numerical simulations the radius $R$ of
the spherical cocoon and the major radius of the torus after the
passage of the shock are approximately equal. Since the major ($R$)
and minor ($r$) radii of the torus can be read off approximately from
a sensitive high-resolution radio map, the compression of the radio
plasma by the shock can be estimated. In the idealised case of an
initially spherical and finally toroidal radio cocoon, the compression
factor is given by

\begin{equation}
C = \frac{V_{\rm sphere}}{V_{\rm torus}} = \frac{4\, \pi\, R^3/3}{2\,
\pi^2\, R\,r^2} = \frac{2\, R^2}{3\,\pi r^2}\,.
\end{equation}
  Assuming the
radio cocoon to be in pressure equilibrium with its environment before
and after the shock passage, the pressure jump in the shock is given
by ${P_2}/{P_1} = C^{\gamma_{\rm rp}}$. If the adiabatic index of the
radio plasma $\gamma_{\rm rp}$ is assumed to be known, e.g
$\gamma_{\rm rp}=4/3$ for an ultra-relativistic equation of state, the
shock strength can be estimated. But even if this assumption is not
justified and the geometry deviates from the idealised geometries
assumed here, the strength of the shock wave should be correlated to
the ratio of the global diameter of a toroidal relic and the thickness
of its filaments. Unfortunately, the quality of the best current radio
maps of relics do not yet allow a quantitative comparison of the shock
strength by comparing the $R/r$ ratios of toroidal relics. But these
maps demonstrate that the necessary sensitivity and resolution might
be reached soon.

If, furthermore, the strength of the shock wave of well resolved
cluster radio relics can be estimated independently from X-ray maps of
the IGM, it would be possible to directly measure the adiabatic index
of radio plasma. This can be done by estimating the slope of a
$\log(P_2/P_1)$ versus $\log(R/r)$ plot for a sample of well observed
radio relics and their shock waves:
\begin{equation}
\log(P_2/P_1) = 2\,\gamma_{\rm rp}\,\log(R/r) - \log(3\,\pi\,\varepsilon/2)\,.
\end{equation}
Here $\varepsilon$ parametrises the influence of the deviation from
the above assumed ideal geometry. The values of $\varepsilon$ should
scatter from relic to relic, but any systematic correlation of
$\varepsilon$ with the shock strength should be weak. Even though the
present radio and X-ray data do not have the required accuracy yet,
at some point in the future this method may enable us to
measure the unknown equation of state of radio plasma.

In our simulations the adiabatic index of the simulated fluid
($\gamma_{\rm rp,sim} =\gamma_{\rm gas} = 5/3$) and the shock strength are
known ($P_2/P_1 = 3.5$ and $P_2/P_1 = 17.4$ for the shock compression
factor $C_{\rm shock}=2$ and $C_{\rm shock}=3.3$ respectively). Thus
we find that the ratio of the length scales $R/r \approx 3$ for
$C_{\rm shock}=2$ and $R/r \approx 5$ for $C_{\rm shock}=3.3$. This is
roughly consistent with our synthetic radio maps. At least the
qualitative correlation of shock strength and diameter ratio is
clearly observed, as can be seen in Figs. \ref{fig:map.WW+70-00},
\ref{fig:map.SW70-00}, and \ref{fig:map.SS+70-45}. Even in the
non-toroidal, filamentary case {\it wSsB} the ratio of the typical
filament diameter to the relic diameter seems to be similar to that in
the toroidal case, as a comparison of Fig. \ref{fig:map.WS60-00} with
Fig. \ref{fig:map.WW+70-00} shows.

\section{Conclusion}

We have presented 3-D MHD simulations of a hot, magnetised bubble that
traverses a shock wave in a much colder and denser environment. This
is assumed to be a fair model for a blob of radio plasma in the IGM
which is passed by a cluster merger shock wave. We have calculated
radio polarisation maps for the relativistic electron population and
computed the spectrum subject to {synchrotron-,} inverse Compton- and
adiabatic energy losses and gains. These maps show that the shock wave
produces filamentary radio emitting structures and, in many cases,
toroidal structures. Such filaments and tori appear to be
observed by very recent high-resolution radio maps of cluster radio
relics (Slee et al. 2001\nocite{2001AJ....122.1172S}). Our simulations
find polarisation patterns which indicate that the magnetic fields are
mostly aligned with the direction of the filaments. This also seems to
be the case for the observed cluster radio relics.

We argued that the formation of filaments and tori is a generic
feature of a shock processed hot bubble of plasma whose
internal sound speed well above the shock speed. We expect this to be
a robust result that will also be found in more realistic simulations,
e.g.  with the proper equation of state, or more realistic initial
magnetic field configurations.

Therefore, we conclude that we have found support for the
hypothesis that cluster radio relics indeed consist of fossil radio
plasma that has been compressed adiabatically by a shock wave, as
proposed by En{\ss}lin \& Gopal-Krishna
(2001)\nocite{2001A&A...366...26E}. Thus, the historical name {\it
`cluster radio relic'} seems to be, accidentally or intuitively, an
appropriate choice.

From our simulations several predictions about the properties of
cluster radio relics can be made -- if our scenario is correct :

The formation of the tori and filaments is not instantaneous. First,
the simulations show a phase in which the radio plasma is strongly
flattened by the shock. During this phase a sheet-like radio relic
with a flat spectrum is observed. Later, the radio plasma moves
towards the edges of this sheet and finally becomes a torus (or a more
complicated, filamentary structure). Since spectral ageing is likely
to have affected these later stages, we expect that the filamentary
relics have a steeper, more bent radio spectrum than the sheet-like
ones.

Our simulations indicate that the diameter $D$ of the bubble of radio
plasma remains approximately constant during the passage of the
shock. It was also found that the final structure consists of radio
filaments of small diameter $d$ that are distributed (often in form of
a torus) in an area of diameter $D$. The compression factor of the
radio plasma is proportional to $(D/d)^2$, and this ratio is a measure
of the shock strength. Thus, the approximate compression factor of the
radio plasma can in principle be read off a sensitive
high-resolution radio map. If this number can be estimated for a
sufficient number of late-stage relics, and if the pressure jump
$P_2/P_1$ of the shock wave can be obtained from detailed X-ray
observations, the unknown adiabatic index $\gamma_{\rm rp}$ of the
radio plasma can be computed using the relation:
\begin{equation}
\frac{P_2}{P_1} \propto \left( \frac{D}{d} \right)^{2\gamma_{\rm rp}} \, .
\end{equation}

The local radio polarisation reflects the complicated
magnetic field structures. But for sufficiently turbulent initial
magnetic fields (as in our simulations) the total integrated
polarisation of a relic reveals the 3-dimensional orientation of the
shock wave. Since the compression aligns the fields with the shock
plane, the sky-projected field distribution is aligned with the
intersection of the shock plane and the sky plane. Thus, the direction
of the total E-polarisation vector yields the sky-projected normal of
the shock wave. In principle the angle between the normal of the shock
and the plane of the sky can be estimated from the fractional
polarisation of the integrated flux. Our simulations indicate that the
viewing angle dependence is similar to that which one would expect if
the magnetic fields were embedded in the external gas and
passively compressed by a shock, even if this may not be the
physical mechanism. But this statement should be treated with some
caution since the simulated polarisation may be sensitive to artifacts
such as the high numerical magnetic diffusivity and the incorrect
adiabatic index of the simulated radio plasma ($\gamma_{\rm rp, sim}=
5/3$ instead of the more likely $\gamma_{\rm rp}= 4/3$). Furthermore,
any initial global field orientation may survive the compression and
distort this relation between viewing angle and polarisation
strength. However, even in this case one expects still a statistical
correlation between these quantities. These points need further
investigation.

We conclude that this work supports the hypothesis that
cluster radio relics are revived bubbles of fossil radio plasma, the so
called {\it radio ghosts}. Spectral aging arguments (En{\ss}lin \&
Gopal-Krishna 2001\nocite{2001A&A...366...26E}) predict the existence
of a sizable population of yet undetected cluster radio relics which
are only observable with sensitive low frequency radio telescopes.

\section*{Acknowledgements}
We thank O.B. Slee, A.L. Roy, M. Murgia, H. Andernach, M. Ehle for
providing us with their observational data prior to publication, and
allowing us to display their 1.4 GHz map of the relic in Abell 85. We
also thank G. Giovannini and L. Feretti for access to their 330 MHz
data of the same relic. We acknowledge A. Kercek's contributions to a
very early stage of this simulation project and E. M{\"u}ller's
comments on the manuscript. Some of the computations reported here
were performed using the UK Astrophysical Fluids Facility
(UKAFF). This work was supported by the European Community Research
and Training Network `The Physics of the Intergalactic Medium'.

\bibliographystyle{plain}


\label{lastpage}

\end{document}